\documentclass[conference]{IEEEtran}
\IEEEoverridecommandlockouts

\usepackage[colorlinks,allcolors=blue]{hyperref}
\usepackage{cite}
\usepackage{amsmath,amssymb,amsfonts}
\usepackage{algorithmic}
\usepackage{graphicx}
\usepackage{textcomp}
\usepackage{xcolor}
\def\BibTeX{{\rm B\kern-.05em{\sc i\kern-.025em b}\kern-.08em
    T\kern-.1667em\lower.7ex\hbox{E}\kern-.125emX}}

\usepackage{tabularx}

\usepackage{multirow}
\usepackage{xspace}
\usepackage{todonotes}
\usepackage{booktabs}

\usepackage{url}

\newcommand\acronym{$\mathsf{EVSOAR}$\xspace}

\makeatletter 
\newcommand{\linebreakand}{%
  \end{@IEEEauthor0halign}
  \hfill\mbox{}\par
  \mbox{}\hfill\begin{@IEEEauthorhalign}
}
\makeatother 

\begin{document}

\title{\acronym: Security Orchestration, Automation and Response via EV Charging Stations}

\author{
\IEEEauthorblockN{Tadeu Freitas\IEEEauthorrefmark{1}
Erick Silva\IEEEauthorrefmark{2},
Rehana Yasmin\IEEEauthorrefmark{3}, 
Ali Shoker\IEEEauthorrefmark{4},\\
Manuel E. Correia\IEEEauthorrefmark{5},
Rolando Martins\IEEEauthorrefmark{6},
Paulo Esteves-Verissimo\IEEEauthorrefmark{7}}
\IEEEauthorblockA{
King Abdullah University of Science and Technology (KAUST) and\\Faculty of Science of University of Porto (FCUP),\\
Email:\{erick.silva\IEEEauthorrefmark{2}, 
rehana.yasmin\IEEEauthorrefmark{3}, 
ali.shoker\IEEEauthorrefmark{4}, 
paulo.verissimo\IEEEauthorrefmark{7}\}@kaust.edu.sa,\\
Email:\{tadeufreitas\IEEEauthorrefmark{1}, 
mdcorrei\IEEEauthorrefmark{5}, 
rmartins\IEEEauthorrefmark{6}\}@fc.up.pt
}
}
\maketitle

\begin{abstract}
Vehicle cybersecurity has emerged as a critical concern, driven by the innovation in the automotive industry, e.g., automomous, electric, or connnected vehicles.
Current efforts to address these challenges are constrained by the limited computational resources of vehicles and the reliance on connected infrastructures.
This motivated the foundation of Vehicle Security Operations Centers (VSOCs) that extend IT-based Security Operations Centers (SOCs) to cover the entire automotive ecosystem, both the in-vehicle and off-vehicle scopes. 
Security Orchestration, Automation, and Response (SOAR) tools are considered key for impelementing an effective cybersecurity solution.
However, existing state-of-the-art solutions depend on infrastructure networks such as 4G, 5G, and WiFi, which often face scalability and congestion issues.

To address these limitations, we propose a novel SOAR architecture \acronym that leverages the EV charging stations for connectivity and computing to enhance vehicle cybersecurity. 
Our EV-specific SOAR architecture enables real-time analysis and automated responses to cybersecurity threats closer to the EV, reducing the cellular latency, bandwidth, and interference limitations.
Our experimental results demonstrate a significant improvement in  latency, stability, and scalability through the infrastructure and the capacity to deploy computationally intensive applications, that are otherwise infeasible within the resource constraints of individual vehicles.
\end{abstract}

\begin{IEEEkeywords}
EVSOAR, vehicle security, cybersecurity, VSOC
\end{IEEEkeywords}

\section{Introduction}
\label{sec:intro}
The widespread adoption of modern Electric Vehicles (EVs) has accelerated the integration of novel technologies, resulting in the evolution of E/E architectures and enabling connectivity to the Internet of Everything. 
This development advanced the concept of the Software-Defined Vehicle~\cite{liu2022impact}. 
Alongside these advancements, a new infrastructure—the EV Charging Station Network (CSN)—was established to support these vehicles' operation and charging needs. 

Although this opens opportunities for new applications and user convenience, it expands the attack surface on the entire EV ecosystem.
The connectivity of EVs has introduced new vulnerabilities, exposing them to various cyberattacks and presenting significant cybersecurity challenges.
Due to the inherent resource constraints in vehicles, implementing traditional and new cybersecurity mechanisms poses unique difficulties.
The traditional Security Operation Center (SOC) was redefined to address these challenges as the Vehicle Security Operation Center (VSOC).
The primary goal of VSOCs is to continuously monitor the behavior of connected vehicles, detect faults and intrusion events, and provide effective cybersecurity responses.
Tools like Security Information and Event Management (SIEM) and, more recently, Security Orchestration, Automation, and Response (SOAR) are typically essential in modern SOCs.

Nevertheless, the shared nature of the connectivity infrastructure—used simultaneously by vehicles and other Internet-connected devices—results in delays in applying security updates and transmitting critical information beyond the vehicle~\cite{shoker2023scaiota}. 
This limitation is evident in the architectures of current VSOCs proposed in the state of the art~\cite{barletta2023knowledge, martin2023testbed}.
Additionally, VSOCs introduced opportunities for collaboration among Original Equipment Manufacturers (OEMs), facilitating the exchange of security intelligence and the delivery of timely security responses, such as patches and rollbacks. 
These functions align closely with the capabilities of SIEM and SOAR solutions, which serve as the backbone of VSOCs for detecting and mitigating cyber threats.

In this work, we propose \acronym, the first SOAR architecture tailored for VSOCs in the context of EVs.
\acronym leverages the EV CSN to efficiently manage the bidirectional flow of vulnerability and threat intelligence data and disseminate software updates and security patches. 
The architecture employs the EV CSN as an edge computing platform, enabling the transfer of resource-intensive logs via direct \textbf{Power Line Communication (PLC)} links. 
This approach reduces reliance on overloaded and costly cellular networks (e.g., LTE/5G). 
Centralized SOAR computations, typically performed in the cloud~\cite{mir2021implementation}, are redistributed across the EV CSN, enhancing the system's responsiveness by enabling real-time updates to detection and prevention models and Over-the-Air (OTA) updates for security patches and software.

The proposed architecture facilitates cross-vendor data correlation through the EV CSN, enriching threat intelligence capabilities while ensuring secure and private data sharing.
For example, Federated Learning (FL) can be employed to enable collaborative model training without compromising data privacy, as discussed in the subsequent sections.
However, this approach introduces challenges, particularly the need to safeguard the charging station, given its physical accessibility, and secure the vehicle's SOAR agent. 
Although there are solutions to address these concerns, they must be adapted for \acronym and ensure that the overhead remains negligible.


To validate the effectiveness and feasibility of \acronym, we demonstrate three cutting-edge use cases:
\begin{itemize}
    \item Log Sharing and Processing: Facilitating the transfer of data logs from EVs to the cloud center for batch processing, with feedback loops to the EV CSN. From our experiments it was possible to visualize a minimum improvement of 41\%.
    \item Federated Learning Integration: Enabling collaborative data sharing across different OEMs via the EV CSN, with federated insights fed back to individual vehicles. Our experiments show that it achieves an accuracy of 93\%, very close to IDS using traditional ML (97\%).
    \item Threat Response: Efficiently addressing detected threats by triggering software updates, rollbacks, or patches delivered through the EV CSN.
\end{itemize}

The results of the experiments indicate that the use of EV CSN provide better connectivity for \acronym, showing a minimum improvement of 41\% for transmitting lower payloads (less 1KB), than the ones used by state-of-the-art VSOCs, SOC4AS and XL-SOC.

The remainder of this paper is organized as follows. 
Section~\ref{sec:related}: Reviews the state of the art in VSOCs and Intrusion Detection Systems (IDSs).
Section~\ref{sec:arch}: Provides a detailed description of \acronym's architecture.
Section~\ref{sec:eval}: Discusses the experiments conducted using \acronym, comparing its performance against other state-of-the-art SOCs, and analyzes the results.
Section~\ref{sec:conc}: Concludes the study and outlines directions for future research.

\section{Related Works}
\label{sec:related}


Automotive security is becoming increasingly complex due to in-vehicle network connections, advanced ECUs, and V2X systems, which expand the attack surface and outpace the capabilities of tools like IDSs, IPSs, gateways, and firewalls~\cite{rathore2022vehicle}. 
Consequently, this expands the attack surface, making vehicles more vulnerable to malicious adversaries.
While SOCs address security gaps by analyzing data and responding to threats~\cite{burkacky2020cybersecurity}, they face challenges in automotive contexts, such as real-time processing constraints, protocol heterogeneity, and limited in-vehicle resources~\cite{nowdehi2019automotive}. 
IDSs, though effective, lack the system-wide visibility and integrative capabilities of SIEM/SOAR, which coordinate insights across OEMs, vehicles, and suppliers for improved threat response. 
Traditional SOCs also fail to account for the contextual nuances of vehicle-specific behavior.

The adaptation of SOCs into VSOCs has been examined extensively. 
Langer et al.~\cite{langer2019establishing} (using the terms CDC and ACDC instead of SOC and VSOC) and Hofbauer et al.~\cite{hofbauer2023soc} identified the minimum legal requirements and necessary modifications for transitioning SOCs to VSOCs to enhance vehicle security.
Saulaiman et al.~\cite{saulaiman2024developing} advanced this by developing a prototype VSOC and evaluating its SIEM capabilities.
Other significant contributions include Meyer et al.~\cite{meyer2020security}, who explored vehicle-targeted attack scenarios without addressing a VSOC implementation, and Barletta et al.\cite{barletta2023v}, who proposed a VSOC framework to improve response and prevention phases through simulations of Denial-of-Service (DoS) and fuzzing attacks.
Martin et al.~\cite{martin2023testbed} introduced a multi-layered architecture embedding distinct SIEM systems in each layer for structured VSOC deployment.
All solutions use wireless communication for their VSOC designs, integrating IDSs or SIEM/SOAR tools to monitor vehicles. 
However, vehicle communication competes with existing devices, further limiting resources and scalability—critical considerations given the growing number of manufactured vehicles.
\acronym takes a distinct approach by utilizing wired communication within the EV CSN infrastructure, ensuring scalability, resource availability, quality of service, and the capacity for integrating additional SOAR tools.

VSOCs are in early stage development but face challenges such as high costs and limited scalability due to reliance on shared communication infrastructure.
EVSOCs present a scalable alternative by utilizing the EV CSN to connect directly with vehicles, avoiding infrastructure strain.
This enables a one-to-one mapping between vehicles and charging stations, with the flexibility to expand alongside new station deployments.

\section{Architecture}
\label{sec:arch}

This section introduces the \acronym architecture, underlying models, and functional applications that enable effective threat detection, automation, and response. We highlight features of \acronym's applications, detailing how the architecture supports security functions, specifically for Intrusion Detection Systems, Federated Learning, and Response Mechanisms. We first start by presenting the threat model.





\subsection{Threat Model}

\subsubsection{Threats and Assumptions} An adversary may attack a vehicle by \textup{(1)} installing compromise security updates to control specific ECU or entire vehicle through \textit{Man-in-the-Middle} (MITM) or \textit{Spoofing} attacks~\cite{elkhail2021vehicle, mahmood2022systematic}; \textup{(2)} launching a remote attack to render a functionality or the entire vehicle unusable by leveraging on wireless communication (e.g., using the wireless module as in ~\cite{greenberg2015jeep}); or \textup{(3)} exploiting software vulnerabilities (e.g., through a Wi-Fi hotspot as in ~\cite{kim2021cybersecurity}). 
Communication threats concern adversary compromising the confidentiality of personal data for profiling or other purposes, and integrity or availability of the data exchanged. 
The adversary may compromise availability by dropping, delaying, or corrupting logs and responses (e.g., \textit{Endless-data}  and \textit{Mix-and-match} attacks ~\cite{mahmood2022systematic}). 
Another challenge is ensuring a secure chain of trust across the architecture while using EV CSNs as edge devices. 
For this it is assumed that a PKI~\cite{corniani2024pki} or an SSI~\cite{kailus2024self} are set up to secure the chain.
It is assumed that the cloud provider and the OEM servers are highly resilient and secure against malicious adversaries where adversaries can compromise one at a time but not both simultaneously, even in the worst-case scenario. More details on attacks are described in ~\cite{xiong2019study, elkhail2021vehicle, mahmood2022systematic}.

\subsubsection{Adversary's capabilities} The adversary can:
\begin{itemize}
    \item \textit{Compromise the vehicle's ECUs.} Compromised ECUs can intercept, alter, or inject messages and suppress alerts designed to warn about malfunctions or security threats. In addition, compromised ECUs can manipulate other ECUs to prevent warnings or security alerts from being generated. Data generated from sensors can also be tampered with by compromised ECUs, causing them to make incorrect decisions or suppress safety mechanisms.
    
    \item \textit{Intercept channel communications.} This includes the interception of channels established between the various components of the architecture, i.e., off-vehicle (e.g., LTE, 5G, Wi-Fi) and in-vehicle (through an OBD-II port, USB, or other means).

    \item \textit{Compromise the cryptographic keys.} An adversary can compromise the cryptographic keys, e.g., through elevation of privileges or stealthy attacks that allow it to perform spoofing attacks or modify the signed hash digests of logs and events.  
    
\end{itemize}

\subsection{Architecture}

\begin{figure*}
    \centering
    \includegraphics[width=0.8\linewidth]{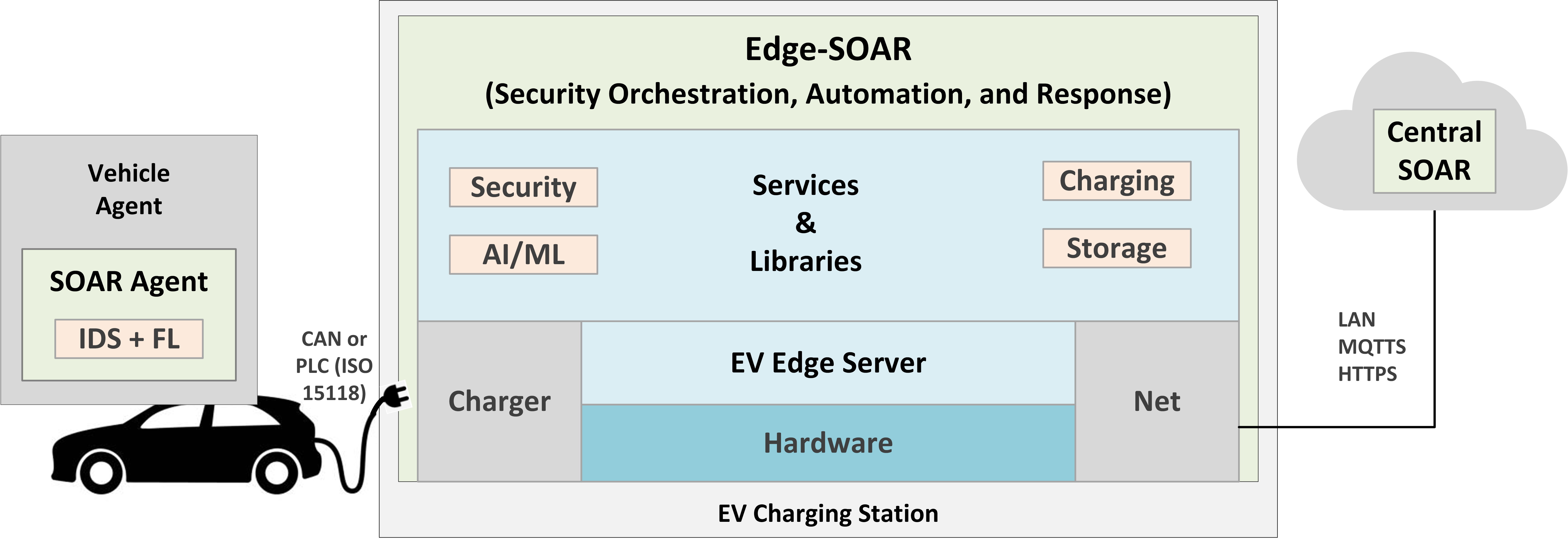}
    \caption{SOAR architecture}
    \label{fig:arch}
\end{figure*}

\begin{figure}[htbp]
    \centering
    \includegraphics[width=0.8\columnwidth]{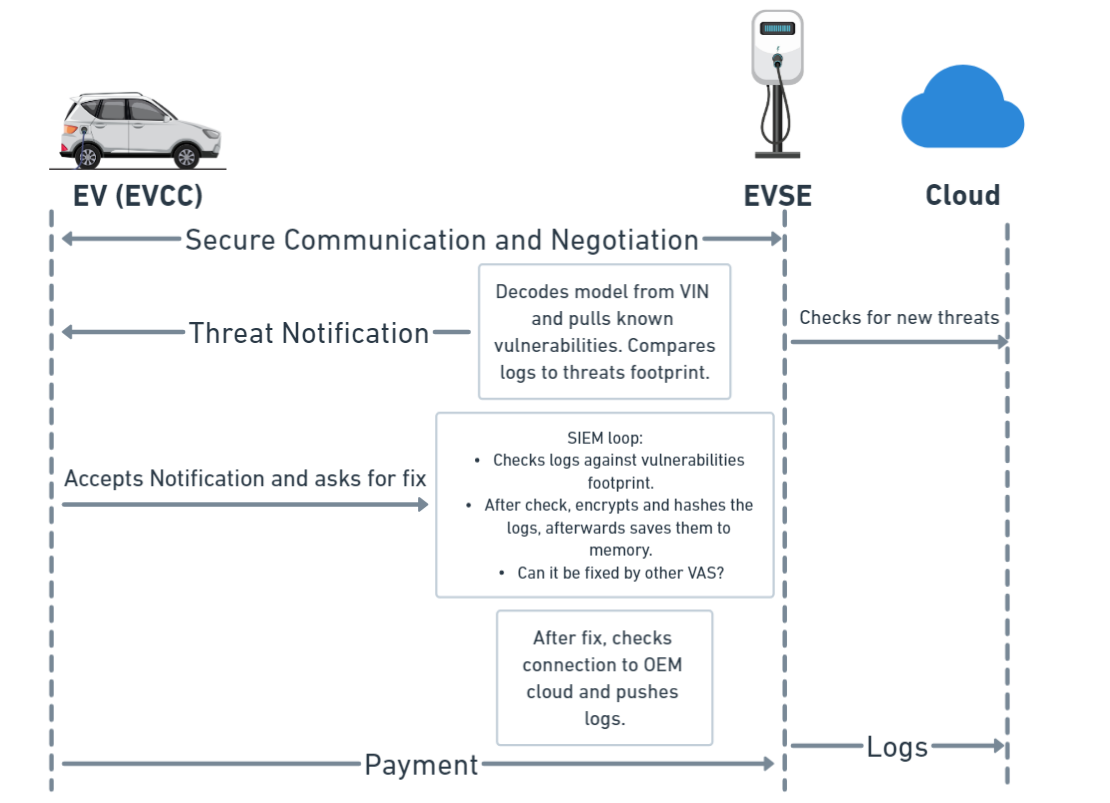}
    \caption{SOAR flowchart}
    \label{fig:flow}
\end{figure}

Figure~\ref{fig:arch} presents the architecture of EV-SOAR. It comprises three main parts: the SOAR Agent installed in vehicle, the Edge-SOAR installed in each charging Point at the EV charging station, and the Central SOAR.
Figure~\ref{fig:flow} illustrates the execution flowchart detailing the process of a vehicle connecting to a charging station and its effects in \acronym.

The \textbf{SOAR agent}, installed within the vehicle, deploys the politics established by the Central SOAR regarding logs, events, and alerts triggering during the vehicle's travels.
Besides this, it can communicate with OEMs to transmit logs/events generated by the circulating vehicles, inform them of possible malfunctions, and request/notify them of the need for a security patch/update through log analysis.
Critical data and alerts may be stored in various locations depending on the vehicle's architecture, including but not limited to the Event Data Recorder, the Central Gateway ECU, the Telematics Control Unit (TCU), the ECU’s internal memory, or the Onboard Diagnostic System.
 
The \textbf{Edge-SOAR}, installed within the charging stations, establishes a physical connection through the charging cable (CCS Type-2 or CHAdeMO 3.0) with the SOAR Agent using secured channels (such as TLS or Virtual Private Network) for data transmission.
Edge-SOAR nodes are interconnected across the EV CSN, facilitating threat intelligence sharing and coordinating responses within a zone or region.
In contrast, stations across broader geographical areas may rely on WAN or internet connections, maintaining resilience through distributed network architecture.
Using the SOAR agent installed within the vehicle, the Edge-SOAR retrieves data from vehicles (logs, events, and alerts) that connect to the charging station through the charging cable.
As such, the charging cable is assumed to include a dedicated software communication line, either bundled with the power cables or utilizing Power-Line Communication (PLC)~\cite{ferreira2011power}.
The logs can be relayed to an Intrusion Detection System (IDS), which analyzes them to detect possible intrusions through pattern analysis and/or sends them to the centralized SOAR for further analysis.

After detecting a positive intrusion, an IDS can trigger an automatic response if the cause is known or inform the central SOAR of the intrusion to assess it and update the automatic responses.
In addition, the Edge-SOAR has a rule-based log assessment that sets off automatic responses to mitigate the cause of the log/event, i.e., malicious event, intrusion, or malfunction.

The \textbf{Central SOAR}, managed by a third party, is connected to all deployed Edge-SOAR systems installed in the EV charging stations participating in the architecture.
Equipped with a Security Operations Team, the Central SOAR gathers all information, preprocesses it, and presents it to the team for further analysis and inspection.
The objective of the centralized SOAR is to analyze events and logs to detect possible security incidents, provide automatic responses that are deployed in the Edge-SOAR, support IDSs by training the model with the retrieved logs (traditional Machine Learning), or aggregate the params received from deployed SOAR Agent's to update the model (Federated Learning).
It also includes two-way communication to exchange security information with the OEM servers.
This allows data transmission on new attack vectors identified by analysts and the reception of security patches or mitigation responses to address these threats. 
The connection is established with the OEM's servers and the suppliers' servers responsible for the deployed software, as the vehicle may run software developed by the OEM or external suppliers.

\subsection{Applications and Features}
Implementing a SOAR-type architecture within the EV CSN introduces new opportunities for security applications and features to enhance the safety of connected vehicles.
Utilizing the physical connection in EVs reduces the data transfer load for security purposes, which current wireless-based solutions typically face, and simplifies the assumptions and policies needed for secure operations. 
This section addresses two applications and a feature enabled by the presented architecture: the application of an IDS using traditional ML (ML IDS) within the charging station, the deployment of an IDS using Federated Learning (FL IDS) within the SOAR Agent, and the support of Automated Response for security notifications and alert events.

\subsubsection{ML IDS} As described in the architecture, when a vehicle connects to a charging station, it uploads the latest security alerts and notification logs generated since its last charging operation to the charging station.
Installing an ML IDS in the charging station enables the evaluation of the logs in search of possible intrusions or malware signs in the vehicle.
This supports an immediate response in scenarios where the threat is known \textit{a priori}.

Since a charging station has limited resources to house a comprehensive IDS, i.e., computation capability,  the idea for this service is to divide it into a central training module deployed at the Central SOAR and a distributed assessment module deployed at the SOAR Agent.
The central module receives new data, preprocesses it, and trains the ML model.
Afterward, the newly generated model is distributed among the Edge SOARs that substitute the locally stored model for the new one for data assessment.
After the identification of compromised data, the Edge SOAR, based on the available information, can take actions toward the connected vehicle depending on the identified threat level, including but not limited to ECU or sensor deactivation, security patch application, or update rollback.

A trade-off in using this service is that data privacy and confidentiality may be compromised.
Although there is research on combining IDS with homomorphic encryption~\cite{sgaglione2019privacy}—an approach ensuring data privacy—it remains computationally heavy, even for standard computers with better computation capability. 
Additionally, logs collected from the vehicle may include data considered private by either the vehicle owner or the OEM, raising potential privacy concerns.

\subsubsection{FL IDS} An alternative to address the privacy challenge of traditional ML IDS is FL, which avoids the transmission of private or sensitive data beyond the vehicle.
This application is integrated at the SOAR Agent level.
Rather than sending logs externally, the FL IDS installed in the SOAR Agent analyzes in-vehicle logs for indicators of potentially compromised messages, which may signal malicious activity.
The SOAR Agent then uses the detection results to initiate an automatic response or to store the information about the compromised component for later communication to the Edge SOAR.

When the vehicle connects to the Edge SOAR, it transmits the FL parameters for aggregation at the Central SOAR through the Edge SOAR.
After aggregation with the FL parameters of other vehicles, the updated FL parameters are sent back to the vehicle.
The trade-off of this application is that, while it protects user and OEM data from being exposed, it suffers from lower accuracy due to the challenge of statistical heterogeneity~\cite{challengesFL}.
The variability in vehicle data can negatively impact the model, as it must generalize across non-identically distributed data.

\subsubsection{Automatic Response} 
SOAR systems automate repetitive threat mitigation tasks, enabling analysts to focus on identifying new vulnerabilities. 
Automated responses, triggered locally at Edge SOARs, reduce reliance on the Central SOAR, enhance scalability, and adapt to evolving threats.
These responses, activated through mechanisms like pattern recognition, IDS, FL, or rule sets, execute actions such as applying security patches, isolating compromised components, updating firmware, or rolling back updates to maintain security.
For example, a compromised infotainment system targeting an ECU can be swiftly deactivated by the Edge SOAR, which isolates the threat and applies a security patch to mitigate the intrusion and protect the vehicle network.


\section{Evaluation}
\label{sec:eval}

\subsection{Experimental Setting}

The \acronym prototype was implemented using Golang combined with the ELK stack~\cite{chhajed2015learning}. 
The ELK stack provides SOAR capabilities, enabling monitoring and threat detection. 
To emulate a realistic deployment environment, \acronym was deployed on the Emulab platform~\cite{emulab}, allowing multiple physical machines (4 physical cores with 12 GB of memory) without interference from other applications. Emulab also enabled the emulation of key network parameters, such as bandwidth, latency, and packet loss ratio (PLR), to match diverse real-world conditions. The code was executed on Ubuntu 20.04 bare-metal machines.

The prototype was tested using five distinct link configurations to represent typical network use cases for vehicle-to-infrastructure and vehicle-to-cloud communications.

\begin{table}[ht]
\caption{Distinct link configurations.}
\centering
\begin{tabular}{|l|c|c|c|}
\hline
\textbf{Type of}  & \textbf{Bandwidth} & \textbf{Latency} & \textbf{PLR} \\ 
\textbf{connection}  & \textbf{(Mbps)} & \textbf{(ms)} & \textbf{(\%)} \\ \hline
\textbf{VSOC-4G}       & 30      & 36        & 0.2       \\ \hline
\textbf{VSOC-5G}    & 100      & 17        & 0.2       \\ \hline
\textbf{RSU-WiFi}  & 27      & 100        & 17      \\ \hline
\textbf{EVSOAR-PLC10M} & 10      & 2        & 0       \\ \hline
\textbf{EVSOAR-PLC100M}        &    100        & 2 & 0 \\ \hline
\textbf{EVSOAR-PLC1G} & 1000      & 2        & 0       \\ \hline
\textbf{Cloud Connectivity}        &    10000    & 0.3 & 0.1 \\ \hline
\end{tabular}
\label{tab:LinkConfig}
\end{table}

\subsection{Evaluation Methodology}
The evaluation compares the state-of-the-art SOC4AS~\cite{barletta2023v} and XL-SOC~\cite{martin2023testbed}, which include SOAR capabilities, with \acronym.
These systems utilize 5G/4G LTE, RSU with WiFi, or physical connection.
The experiments were designed to vary metrics such as bandwidth, latency, and PLR according to the protocol. 
This approach assessed each solution's throughput, network stability, and median RTT.
These metrics were chosen because they highlight the network's performance in real-time applications, particularly for downloading and uploading data to the vehicle.
The collected results also serve as the basis for a practical scenario where a vehicle and a SOC using the different types of connection established by each architecture transmit an intrusion detection and its expected response. 
Additionally, this work includes an experimental evaluation of two distinct scenarios to analyze trade-offs: (1) comparing Machine Learning (ML)-based Intrusion Detection Systems (ML IDS) with FL-based Intrusion Detection Systems (FL IDS) and (2) assessing the performance of FL IDS when using parameters from a single OEM versus the collective pooling of parameters from multiple OEMs.

\subsection{Latency and Throughput}
\begin{figure}
    \centering
    \includegraphics[width=\linewidth]{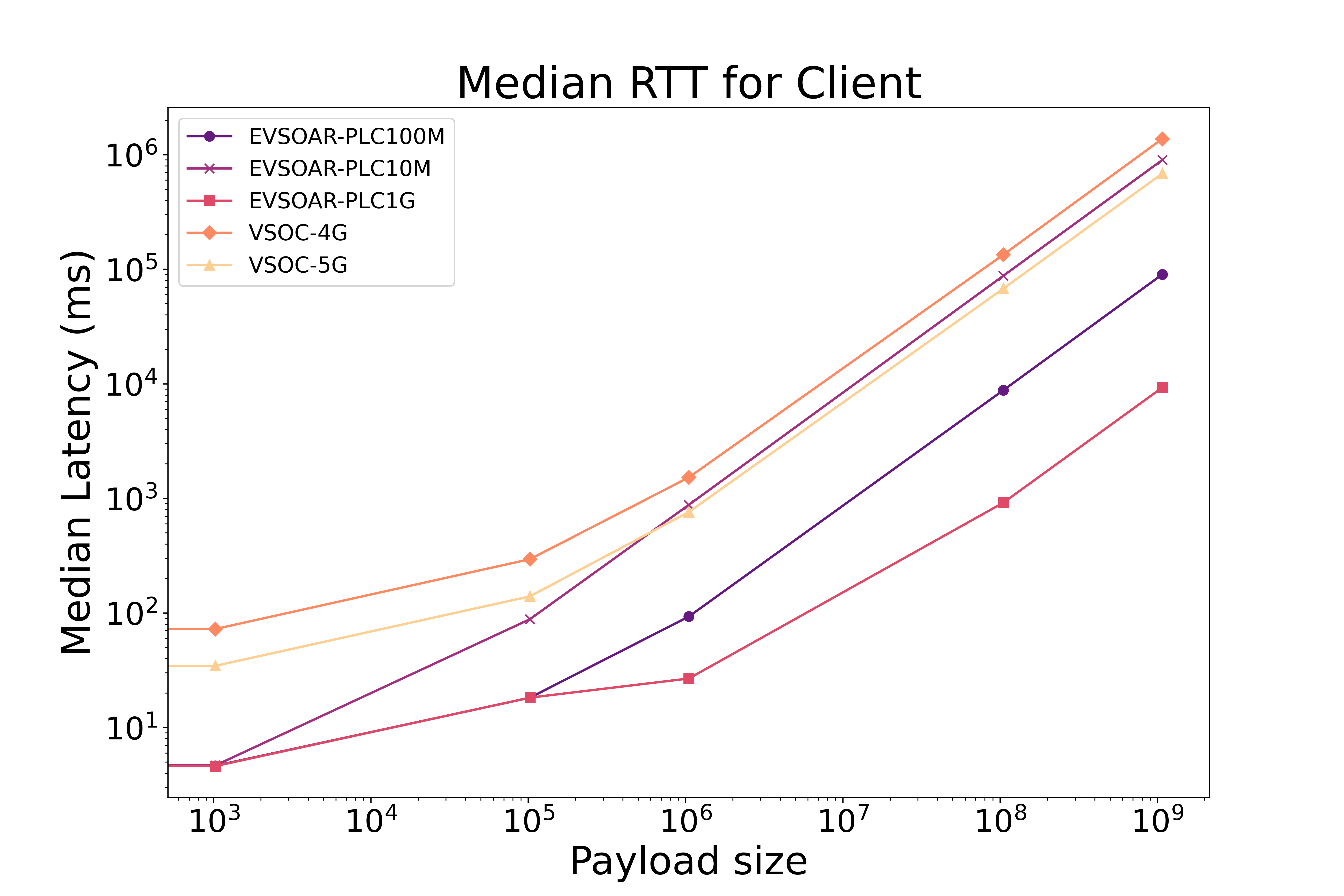}
    \caption{Median Client RTT.}
    \label{fig:rtt}
\end{figure}

\begin{figure}
    \centering
    \includegraphics[width=\linewidth]{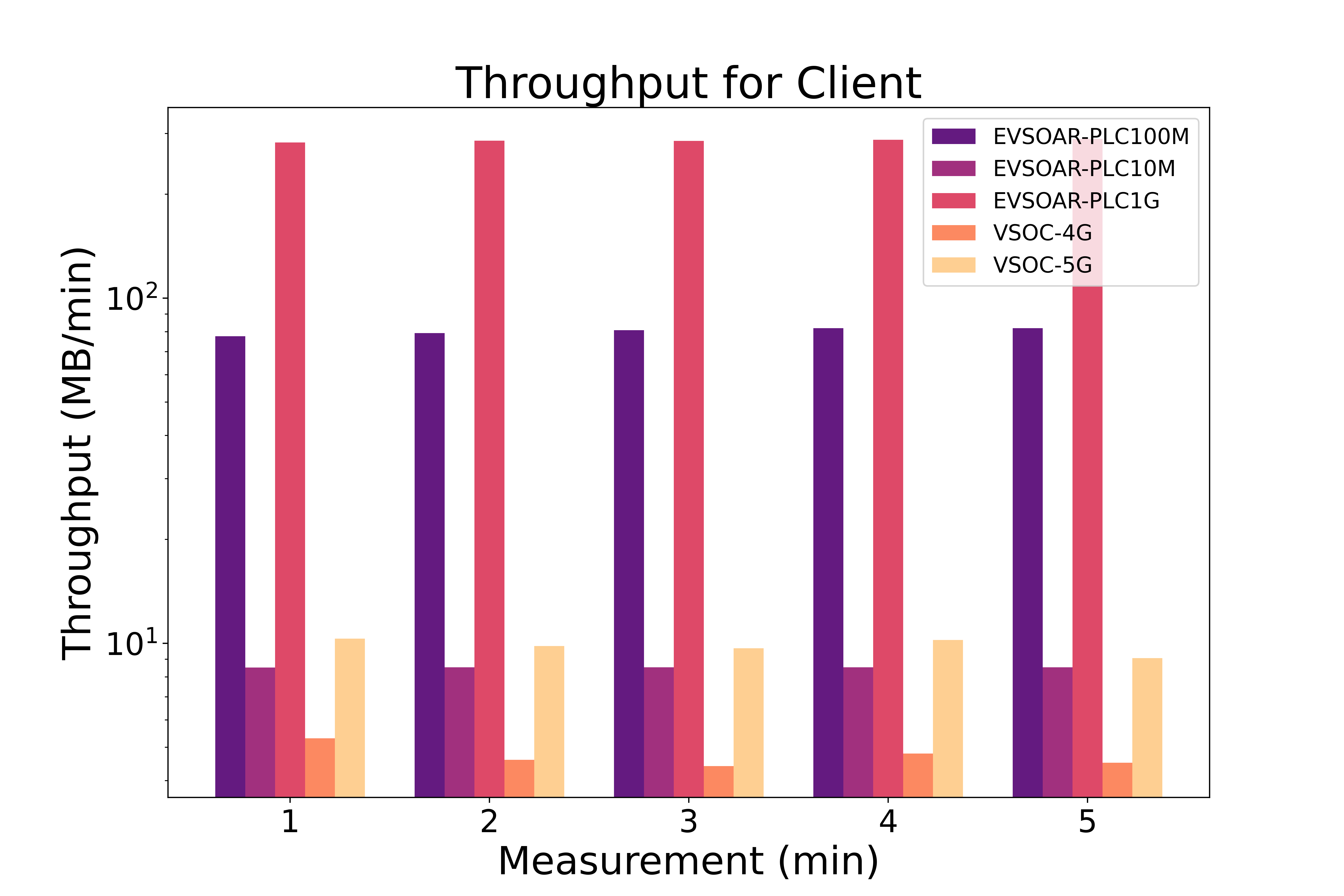}
    \caption{Throughput for Client}
    \label{fig:through}
\end{figure}

Figures~\ref{fig:rtt} and~\ref{fig:through} illustrate the median Round-Trip Time (RTT) and throughput for request and reply operations, respectively.
This experiment evaluates the trade-offs between different architectures, specifically SOC4AS, XL-SOC, and the proposed \acronym.
While the measurements presented in Figure~\ref{fig:rtt} capture the RTT, an approximation to one-way latency is feasible due to the symmetrical connection between the client and server.
Additionally, the response payload size was 0 bytes, making its contribution to the overall RTT almost negligible.
The evaluation reflects a practical scenario involving the transmission of intrusion detection data and the corresponding responses, adhering to the patterns depicted in the figures. 
This scenario accounts for communication flows from both the VSOC and the EVSOC to the vehicle, providing a comprehensive performance assessment of the architectures.
Figures depicting the response transmitted from the server to the client are not included, as the response pattern mirrors the evaluation conducted for the client's transmissions.

By analyzing the figures, it is evident that a wired connection consistently provides higher throughput and lower latency compared to the wireless alternatives. 
This outcome is expected, as wired connections can fully allocate available bandwidth to the vehicle while packet loss remains minimal due to the absence of environmental interference.
By observing the figures, it is possible to visualize that a wired connection consistently delivers higher throughput and lower latency than wireless alternatives.

For smaller payload sizes, \acronym demonstrates a minimum improvement of 41\% over the VSOC-5G architecture, with performance gains increasing as payload size decreases.
This result is unexpected, as the theoretical capacity of VSOC-5G should closely align with that of EVSOAR-PLC1G, a proposed future implementation based on PLC cables.
From our results, EVSOAR-PLC1G consistently takes approximately 10 ms to upload packets of the same size, leveraging its 1 Gbps bandwidth. 
This highlights the impact of bandwidth sharing with other devices, which introduces performance overheads and reduces efficiency, particularly in high-traffic scenarios or competing network demands.

The effects of the high latency result in lower throughput, meaning it takes longer to transmit small packets of data compared to using a wired connection.
For larger payloads, such as those exceeding 1 MB, VSOC-5G begins to surpass \acronym in performance, but only by approximately 15\%.
While this performance gain continues to increase gradually for larger payloads, the growth remains modest. 
It falls short of the theoretical speeds expected for VSOC-5G, demonstrating the effects of shared network environments.


In practical security scenarios, the reliable and consistent throughput offered by wired connections provides a significant advantage, particularly for transferring large volumes of data. 
Wired connections ensure both high speed and efficiency while minimizing delays caused by retransmissions or network congestion, which are common challenges in wireless communication.
This reliability is critical for maintaining timely and secure data exchange in demanding cybersecurity applications.

\subsection{Interference}
\begin{figure}[htbp]
    \centering
    \includegraphics[width=\linewidth]{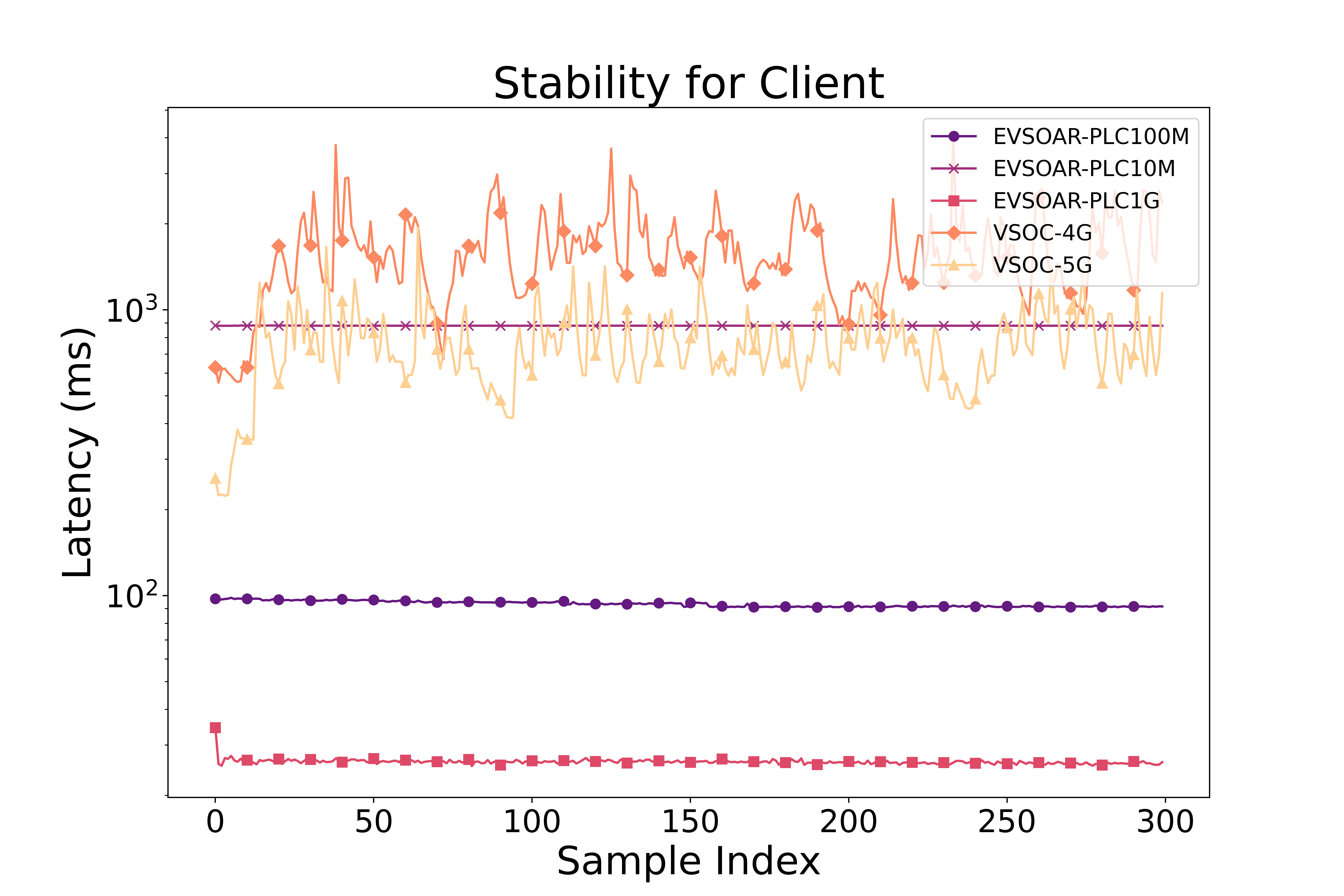}
    \caption{Network stability}
    \label{fig:stability}
\end{figure}

Calculations of interference affecting data transmission were also included to assess impact of network conditions on communication.
These evaluations focused on how packet loss and latency compromise the integrity and performance of transmitted data.
Figure~\ref{fig:stability} depicts the effects of these factors across various protocols for the upstream scenario, which also occurs downstream.
The results underscore the performance of each architecture under different network conditions, offering valuable insights into their suitability for secure data transmission applications.

The observed graphs depict the time required for each packet (transmitted at a rate of one packet per second) to reach its destination. 
The data reveals the inherent instability of wireless connections, such as those used by VSOC-4G and VSOC-5G, compared to the wired connection utilized by \acronym during data transmission between the server and the client. 
This instability negatively affects service quality, particularly in the context of security data transmission.

Wireless networks exhibit significant variability in transmission times, as expected in dynamic environments influenced by vehicle movement and interference from other connected devices and the physical surroundings. 
These variations can range from several hundred milliseconds to thousands of milliseconds, posing challenges for reliable security data transmission, especially for critical server-to-client responses, which can be observed in Figure~\ref{fig:stability}.
In contrast, the wired connectivity employed by \acronym demonstrates much lower variability, with fluctuations limited to at most tens of milliseconds.
This stability makes wired connections far more suitable for the secure and efficient transmission of critical data, ensuring minimal delays and consistent performance.

Combining the presented information with the data illustrated in Figures~\ref{fig:rtt} and~\ref{fig:through}, it can be inferred that relying on wireless networks for security-related requests and responses is not suitable.
This limitation, however, can be effectively addressed by adopting the EV CSN, which offers a more stable and reliable communication channel for such critical operations.



Wired communication is a superior alternative, particularly for VSOC implementations.
This is exemplified by the integration of \acronym with the EV CSN. 
Its inherent stability and consistency in upload and download operations make wired communication highly reliable, ensuring optimal performance for critical and data-sensitive applications.

\subsection{Evaluation Case Studies of EVSOAR Applications}

To evaluate the effectiveness of the \acronym, two applications were developed, each designed to address specific scenarios. 
The first analyzes logs that can be processed openly, utilizing a traditional Intrusion Detection System (IDS) with ML capabilities deployed within the Edge-SOAR. 
The second application focuses on sensitive logs that require privacy, leveraging FL IDS at the SOAR Agent.

\subsubsection{Traditional ML based IDS vs FL based IDS}
This scenario examines the trade-offs between privacy-preserving federated learning and traditional machine learning. 
Federated learning transmits only extracted features, whereas traditional ML requires complete access to raw logs for training.
The key metrics from the ML IDS and the FL IDS were collected and presented in Table~\ref{tab:combinedResults}, demonstrating the trade-offs between these approaches.

\begin{table}[htbp]
\centering
\caption{Combined Results Across ML, FL, and FL Mix}
\label{tab:combinedResults}
\begin{tabular}{|l|l|c|c|c|c|}
\hline
\textbf{Setup} & \textbf{Class} & \textbf{Recall} & \textbf{Support} & \textbf{Data Size} & \textbf{Model} \\
\hline
 & 0 & 1.00 & {\small 2,443,833} & 171 MB  & XGBoost \\
 ML & 1 & 0.82 & {\small 510,341} & (Logs) &  \\
 & \textbf{Acc.} & \multicolumn{2}{|c|}{0.97} &  &  \\
\hline
 FL IDS & 0 & 0.93 & 301,818 & 844 KB  & Feed Forward  \\
 Single & 1 & 0.75 & 45,527 & (Params) & Neural Net. \\
 OEM & \textbf{Acc.} & \multicolumn{2}{|c|}{0.91} &  &  \\
\hline
 FL IDS & 0 & 0.89 & 301,818 & 844 KB  & Feed Forward \\
 Mix & 1 & 0.87 & 45,527 & (Params) &  Neural Net. \\
 OEM & \textbf{Acc.} & \multicolumn{2}{|c|}{0.89} &  &  \\
\hline
\end{tabular}
\end{table}



As shown in Table~\ref{tab:combinedResults}, the ML-based IDS achieves higher accuracy and recall compared to the FL-based IDS. 
This outcome is expected, as the ML approach has complete access to the vehicle-generated logs.
This comprehensive access allows the model to learn the patterns of compromised data (Class 1) effectively, albeit at the cost of requiring hundreds to thousands of megabytes of data. 
Consequently, the ML IDS demonstrates superior classification performance for normal data (Class 0).

In contrast, the FL IDS, particularly with the mixed-OEM approach, achieves a slightly lower accuracy of 89\% compared to the ML IDS.
While this result is expected due to the limited information available in federated learning, it is noteworthy, given the inherent constraints of the FL approach.
Interestingly, the FL IDS performs better in identifying compromised data (Class 1), with a 5\% improvement in recall compared to ML. 
This unexpected result can be attributed to the reduced volume of data the FL model needs to analyze, which may allow it to focus more effectively on identifying minority-class patterns.
However, this improvement comes at the cost of a higher misclassification rate for normal data (89\% accuracy for Class 0). From an application perspective, this trade-off is acceptable, as false positives (misclassifying normal data) are generally less critical than false negatives (failing to identify compromised data). 
Furthermore, the FL IDS preserves privacy by transmitting only federated parameters, ensuring that raw logs remain private and secure.



Based on these findings, the selection between the two approaches depends on the specific requirements of the use case, particularly the importance of privacy.
The ML-based IDS is suitable for analyzing data generated by ECUs, sensors, or other components that do not involve sensitive or private information.
In contrast, the FL-based IDS is better suited for telematics, infotainment, and other elements that monitor vehicle behavior, ensuring that sensitive data remains private.

\subsubsection{FL: Single-OEM data vs. Multi-OEM data}
This scenario explores the trade-offs between using federated learning with data from a single OEM versus multiple OEMs. 
The analysis focuses on the benefits of broader data diversity in multi-OEM setups compared to the simplicity and control offered by single-OEM configurations. 

Leveraging data from a single OEM ensures that information remains localized to the vehicles, thereby preserving user privacy and OEM confidentiality.
However, this localization reduces data variability, limiting the IDS's ability to generalize and effectively identify intrusions.
This limitation is reflected in the results presented in Table~\ref{tab:combinedResults}, where the Single OEM approach correctly identified 75\% of compromised data, compared to the Mix OEM approach, which achieved a higher detection rate of 87\%, as expected.

The inclusion of parameters from multiple OEMs introduces greater data variability, enhancing the IDS's ability to detect anomalies and improving overall performance. 
However, this approach depends on effective coordination and parameter sharing among OEMs, which introduces challenges. 
Sharing parameters may expose latent information embedded within them~\cite{wang2019beyond}, raising privacy concerns (which is a discussion beyond the scope of this paper).

Addressing these challenges requires trust and assurances that all participating OEMs will uphold privacy and confidentiality during parameter exchanges.
These findings highlight the critical trade-off between cooperation, privacy preservation, and IDS performance.
The best approach for an FL IDS is the Mix OEM due to its capability of better detecting compromised data.

\subsection{RSU Simulation}


While RSUs offer extended coverage and secure installations as used in XL-SOC~\cite{martin2023testbed}, their performance degrades as the network size grows, with resources exponentially divided as more vehicles join~\cite{karunathilake2022survey}.
V2V communication extends coverage but introduces security vulnerabilities and delays, allowing compromised vehicles to manipulate data~\cite{ko2020rsu}.
These trade-offs, particularly in time-sensitive processes requiring real-time operation~\cite{martin2023testbed}, limit RSUs' suitability for critical applications despite their cost-effectiveness and resistance to physical tampering.
\section{Conclusion}
\label{sec:conc}
In this work, we introduced \acronym, a novel Security Orchestration, Automation and Response (SOAR) architecture tailored to enhance electric vehicle cybersecurity. 
State-of-the-art solutions have faced challenges due to their dependence on infrastructure networks that are not designed for vehicle-specific connectivity and scalability. 
Moreover, resource constraints within vehicles often limit the applicability of traditional security solutions.
Our proposed solution, \acronym, leverages the untapped potential of the EV CSN infrastructure to implement an Edge-SOAR. 
The design integrates a Central SOAR, enabling a federated approach for OEMs to collaborate, share security intelligence, and collectively counteract malicious adversaries.
Our experiments with the FL IDS demonstrated this collaborative potential, which showed improved detection performance when OEMs share knowledge.
Using the EV CSN, \acronym effectively addresses the challenges and limitations of prior solutions. 
Our results highlight the advantages of \acronym in ensuring secure and scalable connectivity while maintaining robust cybersecurity measures.



\bibliographystyle{IEEEtran}
\bibliography{ref}

\begin{thebibliography}{10}
\providecommand{\url}[1]{#1}
\csname url@samestyle\endcsname
\providecommand{\newblock}{\relax}
\providecommand{\bibinfo}[2]{#2}
\providecommand{\BIBentrySTDinterwordspacing}{\spaceskip=0pt\relax}
\providecommand{\BIBentryALTinterwordstretchfactor}{4}
\providecommand{\BIBentryALTinterwordspacing}{\spaceskip=\fontdimen2\font plus
\BIBentryALTinterwordstretchfactor\fontdimen3\font minus \fontdimen4\font\relax}
\providecommand{\BIBforeignlanguage}[2]{{%
\expandafter\ifx\csname l@#1\endcsname\relax
\typeout{** WARNING: IEEEtran.bst: No hyphenation pattern has been}%
\typeout{** loaded for the language `#1'. Using the pattern for}%
\typeout{** the default language instead.}%
\else
\language=\csname l@#1\endcsname
\fi
#2}}
\providecommand{\BIBdecl}{\relax}
\BIBdecl

\bibitem{liu2022impact}
Z.~Liu, W.~Zhang, and F.~Zhao, ``Impact, challenges and prospect of software-defined vehicles,'' \emph{Automotive Innovation}, vol.~5, no.~2, pp. 180--194, 2022.

\bibitem{shoker2023scaiota}
A.~Shoker, F.~Alves, and P.~Esteves-Verissimo, ``Scalota: Scalable secure over-the-air software updates for vehicles,'' in \emph{2023 42nd International SRDS}.\hskip 1em plus 0.5em minus 0.4em\relax IEEE, 2023, pp. 151--161.

\bibitem{barletta2023knowledge}
V.~S. Barletta, D.~Caivano, M.~De~Vincentiis, A.~Pal, and F.~Volpe, ``Automotive knowledge base for supporting vehicle-soc analysts,'' in \emph{2023 IEEE International Conference on MetroXRAINE}, 2023, pp. 960--965.

\bibitem{martin2023testbed}
M.~Mart{\'\i}n-P{\'e}rez, J.~M. Parella, J.~Fern{\'a}ndez, P.~de~Juan~Fidalgo, J.~Casademont, A.~{\'A}. Romero, and R.~Diaz-Rodriguez, ``A testbed for a nearby-context aware: Threat detection and mitigation system for connected vehicles,'' in \emph{2023 IEEE JNIC}, 2023, pp. 1--8.

\bibitem{mir2021implementation}
A.~W. Mir and R.~K. Ramachandran, ``Implementation of security orchestration, automation and response (soar) in smart grid-based scada systems,'' in \emph{Sixth ICICA 2020}.\hskip 1em plus 0.5em minus 0.4em\relax Springer, pp. 157--169.

\bibitem{rathore2022vehicle}
R.~S. Rathore, C.~Hewage, O.~Kaiwartya, and J.~Lloret, ``In-vehicle communication cyber security: challenges and solutions,'' \emph{Sensors}, vol.~22, no.~17, p. 6679, 2022.

\bibitem{burkacky2020cybersecurity}
O.~Burkacky \emph{et~al.}, ``Cybersecurity in automotive,'' McKinsey, Tech. Rep., 2020.

\bibitem{nowdehi2019automotive}
N.~Nowdehi, ``Automotive communication security,'' 2019.

\bibitem{langer2019establishing}
F.~Langer, ``Establishing an automotive cyber defense center,'' 2019.

\bibitem{hofbauer2023soc}
J.~Hofbauer, K.~K.~G. Buquerin, and H.-J. Hof, \emph{From SOC to VSOC: Transferring Key Requirements for Efficient Vehicle Security Operations}.\hskip 1em plus 0.5em minus 0.4em\relax Ruhr-Universit{\"a}t Bochum, 2023.

\bibitem{saulaiman2024developing}
M.~N.-E. Saulaiman, B.~L. Iv{\'a}nyi, E.~Kail, T.~G. Pozsonyi, K.~Z. K{\"o}vesi, R.~Kail, B.~M. T{\'o}th, {\'A}.~Csilling, and A.~B{\'a}n{\'a}ti, ``Developing siem and log management for automotive network in a simulated environment,'' in \emph{2024 IEEE 22nd Jubilee International SISY}, pp. 000\,239--000\,244.

\bibitem{meyer2020security}
P.~Meyer, T.~Hackel, F.~Langer, L.~Stahlbock, J.~Decker, S.~A. Eckhardt, F.~Korf, T.~C. Schmidt, and F.~Sch{\"u}ppel, ``A security infrastructure for vehicular information using sdn, intrusion detection, and a defense center in the cloud,'' in \emph{2020 IEEE VNC}, 2020, pp. 1--2.

\bibitem{barletta2023v}
V.~S. Barletta, D.~Caivano, M.~D. Vincentiis, A.~Ragone, M.~Scalera, and M.~{\'A}.~S. Mart{\'\i}n, ``V-soc4as: A vehicle-soc for improving automotive security,'' \emph{Algorithms}, vol.~16, no.~2, p. 112, 2023.

\bibitem{elkhail2021vehicle}
A.~A. Elkhail, R.~U.~D. Refat, R.~Habre, A.~Hafeez, A.~Bacha, and H.~Malik, ``Vehicle security: A survey of security issues and vulnerabilities, malware attacks and defenses,'' \emph{IEEE Access}, vol.~9, pp. 162\,401--162\,437, 2021.

\bibitem{mahmood2022systematic}
S.~Mahmood, H.~N. Nguyen, and S.~A. Shaikh, ``Systematic threat assessment and security testing of automotive over-the-air (ota) updates,'' \emph{Vehicular Communications}, vol.~35, p. 100468, 2022.

\bibitem{greenberg2015jeep}
A.~Greenberg, ``Hackers remotely kill a jeep on the highway—with me in it,'' \emph{Wired}, July 2015, accessed: 2024-11-07.

\bibitem{kim2021cybersecurity}
K.~Kim, J.~S. Kim, S.~Jeong, J.-H. Park, and H.~K. Kim, ``Cybersecurity for autonomous vehicles: Review of attacks and defense,'' \emph{Computers \& security}, vol. 103, p. 102150, 2021.

\bibitem{corniani2024pki}
E.~Corniani and M.~G. Todeschini, ``Pki implementation for the cybersecurity of electric vehicle charging infrastructures,'' in \emph{2024 AEIT International Annual Conference (AEIT)}, 2024, pp. 1--6.

\bibitem{kailus2024self}
A.~Kailus, D.~Kern, and C.~Krau{\ss}, ``Self-sovereign identity for electric vehicle charging,'' in \emph{International Conference on Applied Cryptography and Network Security}.\hskip 1em plus 0.5em minus 0.4em\relax Springer, 2024, pp. 137--162.

\bibitem{xiong2019study}
W.~Xiong, M.~G{\"u}lsever, K.~M. Kaya, and R.~Lagerstr{\"o}m, ``A study of security vulnerabilities and software weaknesses in vehicles,'' in \emph{Secure IT Systems: 24th Nordic Conference, NordSec 2019, Aalborg, Denmark, November 18--20, 2019, Proceedings 24}.\hskip 1em plus 0.5em minus 0.4em\relax Springer, 2019, pp. 204--218.

\bibitem{ferreira2011power}
H.~C. Ferreira, L.~Lampe, J.~Newbury, and T.~G. Swart, \emph{Power line communications: theory and applications for narrowband and broadband communications over power lines}.\hskip 1em plus 0.5em minus 0.4em\relax John Wiley \& Sons, 2011.

\bibitem{sgaglione2019privacy}
L.~Sgaglione, L.~Coppolino, S.~D'Antonio, G.~Mazzeo, L.~Romano, D.~Cotroneo, and A.~Scognamiglio, ``Privacy preserving intrusion detection via homomorphic encryption,'' in \emph{2019 IEEE 28th International Conference on WETICE}.\hskip 1em plus 0.5em minus 0.4em\relax IEEE, 2019, pp. 321--326.

\bibitem{challengesFL}
T.~Li, A.~K. Sahu, A.~Talwalkar, and V.~Smith, ``Federated learning: Challenges, methods, and future directions,'' \emph{IEEE Signal Processing Magazine}, vol.~37, no.~3, pp. 50--60, 2020.

\bibitem{chhajed2015learning}
S.~Chhajed, \emph{Learning ELK stack}.\hskip 1em plus 0.5em minus 0.4em\relax Packt Publishing Ltd, 2015.

\bibitem{emulab}
{Emulab.net}, ``Emulab testbed,'' \url{https://www.emulab.net}.

\bibitem{wang2019beyond}
Z.~Wang, M.~Song, Z.~Zhang, Y.~Song, Q.~Wang, and H.~Qi, ``Beyond inferring class representatives: User-level privacy leakage from federated learning,'' in \emph{IEEE INFOCOM 2019-IEEE conference on computer communications}.\hskip 1em plus 0.5em minus 0.4em\relax IEEE, 2019, pp. 2512--2520.

\bibitem{karunathilake2022survey}
T.~Karunathilake and A.~F{\"o}rster, ``A survey on mobile road side units in vanets,'' \emph{Vehicles}, vol.~4, no.~2, pp. 482--500, 2022.

\bibitem{ko2020rsu}
B.~Ko, K.~Liu, S.~H. Son, and K.-J. Park, ``Rsu-assisted adaptive scheduling for vehicle-to-vehicle data sharing in bidirectional road scenarios,'' \emph{IEEE Transactions on Intelligent Transportation Systems}, vol.~22, no.~2, pp. 977--989, 2020.

\end{thebibliography}

\end{document}